\documentclass[10pt,twocolumn,letterpaper]{article}

\usepackage{iccv}              

\usepackage[dvipsnames]{xcolor}

\definecolor{iccvblue}{rgb}{0.21,0.49,0.74}
\usepackage[pagebackref,breaklinks,colorlinks,allcolors=iccvblue]{hyperref}
\usepackage{pifont, arydshln, svg, pgfplots, multirow, graphicx}

\def\paperID{11077} 
\def\confName{ICCV}
\def\confYear{2025}

\title{AdaptSR: Low-Rank Adaptation for Efficient and Scalable Real-World Super-Resolution}

\author{Cansu Korkmaz\\
Koc University \\
{\tt\small ckorkmaz14@ku.edu.tr}
\and
Nancy Mehta\\
University of W\"urzburg\\
{\tt\small nancy.mehta@uni-wuerzburg.de}
\and
Radu Timofte\\
University of W\"urzburg\\
{\tt\small radu.timofte@uni-wuerzburg.de}
}

\begin{document}

\definecolor{darkgray176}{RGB}{176,176,176}
\definecolor{darkolivegreen1139724}{RGB}{113,97,24}
\definecolor{darkorange2551160}{RGB}{255,116,0}
\definecolor{darkslategray178045}{RGB}{17,80,45}
\definecolor{darkslategray404747}{RGB}{40,47,47}
\definecolor{darkviolet1910191}{RGB}{191,0,191}

\definecolor{darkcyan18130162}{RGB}{18,130,162}
\definecolor{darkgray176}{RGB}{176,176,176}
\definecolor{darkorange2551160}{RGB}{255,116,0}
\definecolor{darkslategray178045}{RGB}{17,80,45}
\definecolor{darkviolet1910191}{RGB}{191,0,191}
\definecolor{mediumpurple159134192}{RGB}{159,134,192}

\maketitle


\begin{abstract}

Recovering high-frequency details and textures from low-resolution images remains a fundamental challenge in super-resolution (SR), especially when real-world degradations are complex and unknown. While GAN-based methods enhance realism, they suffer from training instability and introduce unnatural artifacts. Diffusion models, though promising, demand excessive computational resources, often requiring multiple GPU days—even for single-step variants. Rather than naively fine-tuning entire models or adopting unstable generative approaches, we introduce AdaptSR, a low-rank adaptation (LoRA) framework that efficiently repurposes bicubic-trained SR models for real-world tasks. AdaptSR leverages architecture-specific insights and selective layer updates to optimize real SR adaptation. By updating only lightweight LoRA layers while keeping the pretrained backbone intact, it captures domain-specific adjustments without adding inference cost, as the adapted layers merge seamlessly post-training. This efficient adaptation not only reduces memory and compute requirements but also makes real-world SR feasible on lightweight hardware. Our experiments demonstrate that AdaptSR outperforms GAN and diffusion-based SR methods by up to 4 dB in PSNR and 2$\%$ in perceptual scores on real SR benchmarks. More impressively, it matches or exceeds full model fine-tuning while training 92$\%$ fewer parameters, enabling rapid adaptation to real SR tasks within minutes. \url{github.com/mandalinadagi/AdaptSR}
\end{abstract}

\vspace{-0.5cm}
\section{Introduction}
\label{sec:intro}

\begin{figure}
\centering
\includegraphics[width=\linewidth]{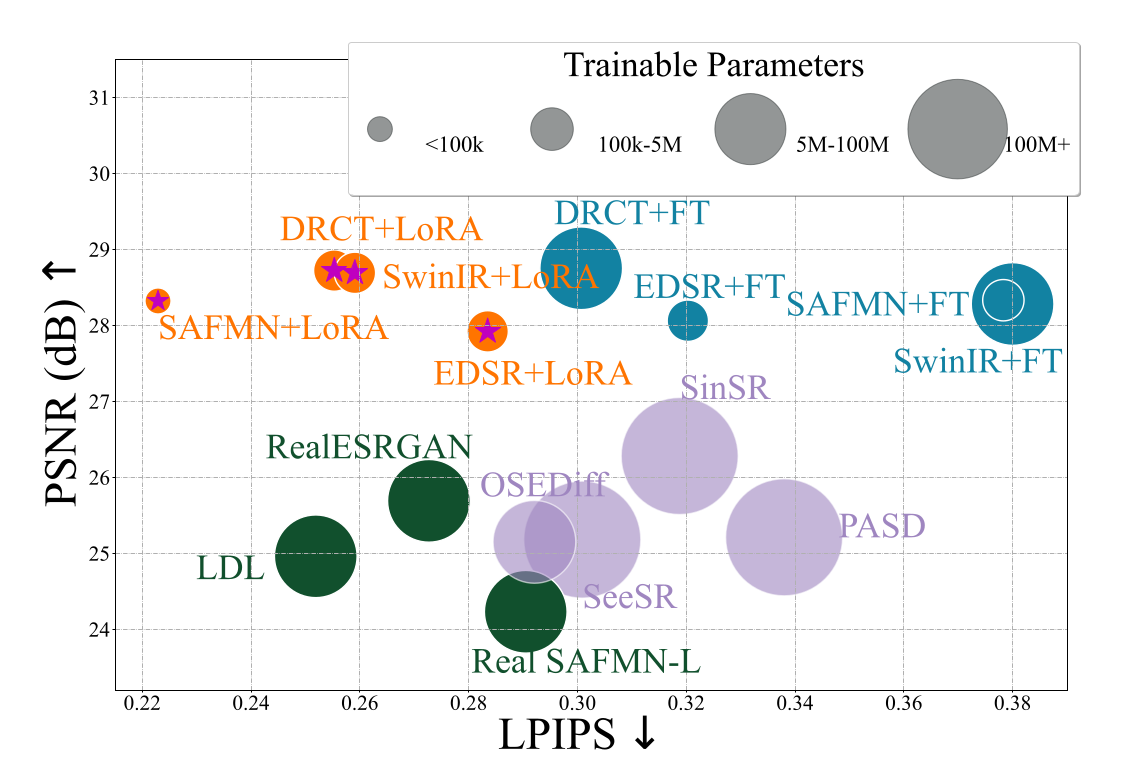}
\caption{Comparison of model complexity and performance between our proposed low-rank adaptation models \textcolor{darkorange2551160}{(orange)} and other real-world SR approaches, including GAN-based \textcolor{darkslategray178045}{(green)} and diffusion-based \textcolor{mediumpurple159134192}{(purple)} models, for $\times$4 SR on the RealSR \cite{realsr_cai2019toward} dataset. Baseline models with full fine-tuning are depicted in \textcolor{darkcyan18130162}{blue}, and our LoRA models for the same baselines achieve better PSNR and LPIPS \cite{lpips} scores with fewer trainable parameters.} \vspace{-0.5cm}
\label{fig:model_complexity_vs_psnr}
\end{figure}


Image super-resolution (SR) aims to reconstruct high-resolution (HR) images from low-resolution (LR) inputs and has essential applications in fields like medical \cite{medical_sr_isaac2015super, medical_sr_qiu2023medical}, optical \cite{optical_lin2024super, optical_rosen2024roadmap}, and satellite imaging \cite{satelite_sr_chen2023large, satelite_sr_karwowska2022using}. While popular convolutional neural network (CNN) and Transformer-based SR models \cite{EDSR2017, RCAN2018, wang2018esrgan, zhang2018residual, niu_han, liang2021swinir, cat_chen2022cross, attention_zhang2022efficient, transformer_Chen_2023_ICCV, chen2023activating, korkmaz2024wavelettention} have advanced architectures, they often rely on simplified degradation models like bicubic downsampling, limiting their generalization to real-world images with complex degradations. Fine-tuning on real-world data helps bridge this gap \cite{wang2018esrgan, liang2022details, wang2024exploiting, ft_yu2024scaling}, but it is computationally intensive, requiring substantial memory and long training times, which poses challenges for scalability and deployment on resource-constrained devices.

Real-world SR (Real SR) \cite{realsr_cai2019toward, drealsr_wei2020component} has gained traction as an alternative to synthetic SR, focusing on explicitly training models for real-world degradations \cite{bsrgan_zhang2021designing, wang2021real, lora_diff_realsr_wu2024one, lora_diff_realsr_wu2024seesr, ZSSR}. GAN-based approaches \cite{ian_gan, wang2018esrgan, wang2021real} leverage texture and structural priors to enhance realism but often suffer from artifacts and instability. Diffusion models (DMs) \cite{kawar2022denoising, ho2020denoising, dhariwal2021diffusion, ldm_rombach2022high, yang2023pasd, wang2024exploiting} achieve state-of-the-art (SoTA) performance by embedding natural image priors, yet their multi-step process is computationally prohibitive. While recent one-step DMs \cite{lora_diff_realsr_wu2024one, lora_diff_realsr_wu2024seesr} improve efficiency, they compromise detail quality. Given the high cost of fine-tuning bicubic-trained models and the limitations of explicitly trained Real SR methods, we pose two key research questions: (1) Can bicubic-trained Transformer/CNN SR models generalize to real-world scenarios without costly fine-tuning? (2)  If so, can they surpass GAN- and diffusion-based Real-SR methods in both generalization and efficiency?

To address the aforementioned issues and bridge the domain gap between synthetic and real-world SR, we introduce AdaptSR, an efficient domain adaptation framework leveraging Low-Rank Adaptation (LoRA) \cite{lora_orig_hu2021lora}. 
Unlike conventional bicubic super-resolution (SR) methods (\textit{e.g.}, SwinIR \cite{liang2021swinir}, SAFMN \cite{safmn_sun2023spatially}), which necessitate comprehensive fine-tuning of all model parameters to adapt to real-world degradation scenarios, AdaptSR employs a parameter-efficient fine-tuning (PEFT) approach. This strategy selectively refines only 8\% of the parameters within bicubic-trained SR models, leading to significant reductions in memory consumption and computational overhead. 
While LoRA \cite{lora_orig_hu2021lora} has been utilized for task adaptation in various domains, AdaptSR is the first to tailor it specifically for real SR, transforming pretrained bicubic models into real-world solutions without full retraining. By enabling rapid adaptation in minutes, even on resource-limited hardware, our approach makes high-quality SR both practical and efficient. Moreover, we provide in-depth analysis of optimal module selection (Sec.~\ref{sec:module-wise}) and a comprehensive comparison with other PEFT methods (Sec.~\ref{ss:comparison_peft}), establishing AdaptSR as a pioneering, lightweight solution for real-world SR.

We validate our method across diverse SR architectures, including the popular CNN-based and transformer-based models ranging from efficient to large-scale designs. 
The proposed framework not only enhances the performance of SoTA synthetically trained CNN and transformer-based SR models but also surpasses GAN and diffusion methods explicitly trained on real-world data. As shown in Figure \ref{fig:model_complexity_vs_psnr}, the (+LoRA) achieves a gain of up to 3 dB in PSNR and a 2$\%$ improvement in perceptual score \cite{lpips} compared to SoTA diffusion model OSEDiff \cite{lora_diff_realsr_wu2024one} while updating only 886k parameters (1$\%$ of OSEDiff). 
To summarize, our primary contributions are:
\noindent
\begin{itemize}
    \item We introduce AdaptSR, a LoRA-based domain adaptation framework that effectively bridges the domain gap between bicubic-trained and real-world SR models, achieving drastic reductions in memory consumption and training time compared to expensive full fine-tuning.
    \item We propose an adaptive merging strategy that seamlessly integrates fine-tuned low-rank updates into the base model, ensuring zero inference overhead while preserving generalization across unseen degradations. Also, we provide a rigorous analysis of optimal layer selection and rank configurations to maximize adaptation efficiency.
    \item We validate the broad applicability of our approach across various SR baselines, including CNNs and Transformers, achieving up to a 4 dB PSNR improvement and a 2$\%$ gain in perceptual score while training only 8$\%$ of model parameters compared to SoTA GAN and diffusion approaches on real-world SR benchmarks. 
\end{itemize}


\section{Related Work}
\label{sec:formatting}
\subsection{Real-World Image Super-Resolution}
Starting with SRCNN \cite{srcnn}, deep learning-based convolutional SR networks \cite{EDSR2017, RCAN2018, wang2018esrgan, zhang2018residual, liang2022efficient, niu_han} achieved notable progress. Later, after the introduction of SwinIR \cite{liang2021swinir}, transformer-based SR networks \cite{cat_chen2022cross, attention_zhang2022efficient, transformer_Chen_2023_ICCV, chen2023activating, korkmaz2024wavelettention} gain popularity by significantly boosting the visual quality of the reconstructed images. However, these models typically assume simple, known degradations (e.g., bicubic downsampling), limiting their performance on real-world images with complex, unknown degradations.

To address real SR challenges, GAN-based models \cite{ian_gan} like SRGAN \cite{ledig2017photorealistic}, BSRGAN \cite{bsrgan_zhang2021designing} and Real-ESRGAN \cite{wang2021real} introduced sophisticated degradation models for real-world data. More recent methods like LDL \cite{liang2022details} and DeSRA \cite{xie2023desra} improve artifact suppression using local statistics, however, GANs still struggle with instability and introduce unnatural artifacts. Recently introduced diffusion models \cite{ho2020denoising, dhariwal2021diffusion, kawar2022denoising, yue2023resshift, ft_yu2024scaling, wang2024sinsr, rebuttal_lin2025diffbir} have also been applied to real SR tasks, such as StableSR \cite{wang2024exploiting} and PASD \cite{yang2023pasd} use feature warping to enhance quality. However, they require hundreds of steps to complete the diffusion process along with excessive memory consumption. Unlike these methods, our approach achieves high-quality SR in real-world settings with minimal training time, enabling rapid adaptation to real SR tasks in minutes rather than hours or days.

\begin{figure*}[t!]
\centering
\includegraphics[width=\linewidth]{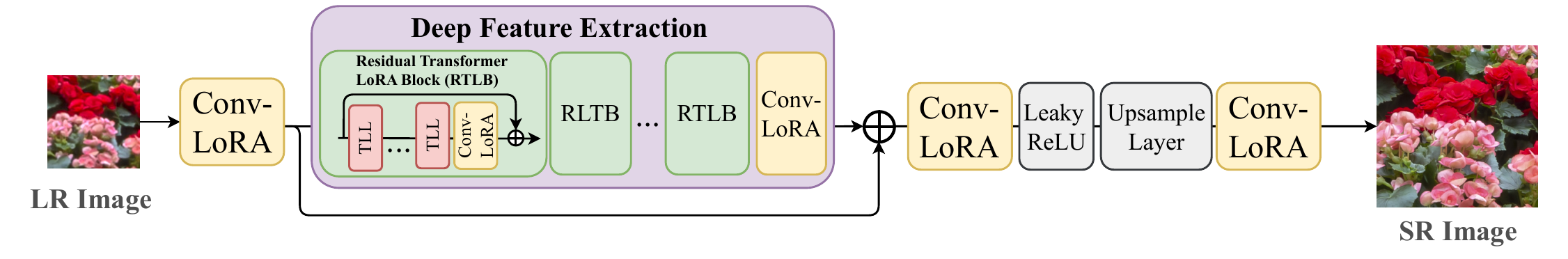}
\caption{An illustration of the proposed AdaptSR, LoRA layers are integrated into the frozen transformer architecture and thus can be seamlessly inserted into various CNN/transformer-based pre-trained SR models to adapt them into real SR. LoRA-modified layers, such as convolutional and attention layers, reduce parameters and computational load, enabling efficient, high-resolution outputs.}
\label{fig:arch}
\end{figure*}

\begin{figure}
\centering
\includegraphics[width=0.9\linewidth]{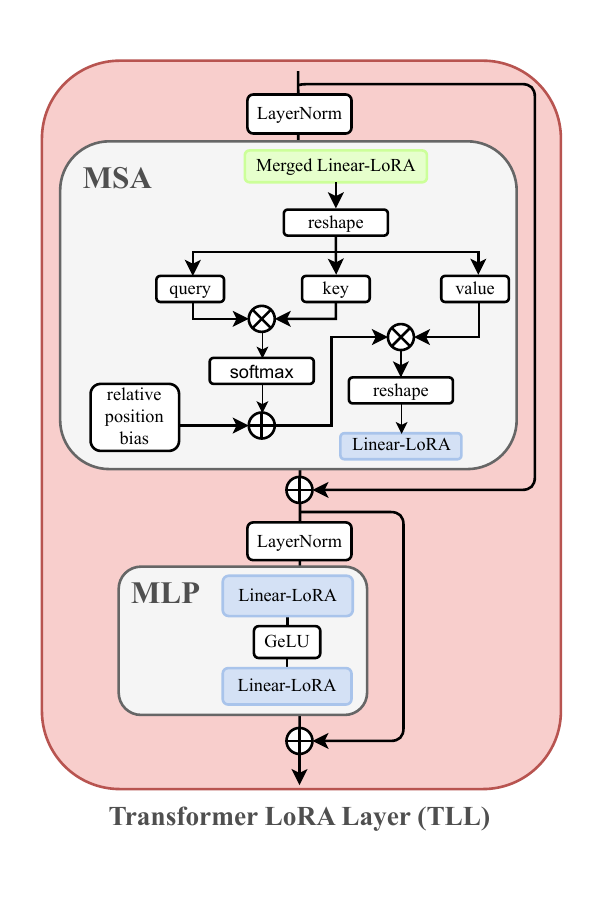}
\vspace{-0.4cm}
\caption{Transformer LoRA Layers (TLL) include LoRA-modified multi-head self-attention (MSA) and multi-layer perceptron (MLP) layers, using Linear-LoRA for efficient domain adaptation with reduced parameters.} \vspace{-0.5cm}
\label{fig:msa}
\end{figure}

\subsection{Parameter Efficient Fine-Tuning for Real SR }

Parameter-efficient fine-tuning (PEFT) \cite{lora_orig_hu2021lora, hu2023llm, lora_llm_dettmers2024qlora, rebuttal_xie2023difffit} methods like adapters \cite{hu2023llm, lei2023conditional, rebuttal_arc_dong2024efficient} and LoRA \cite{lora_orig_hu2021lora, lora_llm_dettmers2024qlora} have become crucial for fine-tuning large pre-trained models with fewer resources. Adapters add small modules between layers for task-specific updates, while LoRA performs low-rank adaptations of model weight matrices, enhancing efficiency without increasing inference-time parameters. LoRA has shown versatility across NLP \cite{lora_llm_hou2023large, lora_llm_dettmers2024qlora}, computer vision \cite{lora_diff_realsr_wu2024seesr, lora_ir_park2024contribution, lora_ir_tian2024instruct, agiza2024mtlora, borse2025foura} and image reconstruction applications \cite{lora_ir_park2024contribution, lora_ir_tian2024instruct} due to its adaptability and computational benefits.

In SR, LoRA has been successfully applied to diffusion-based models, such as SeeSR \cite{lora_diff_realsr_wu2024seesr} and OSEDiff \cite{lora_diff_realsr_wu2024one}, to leverage text-to-image generative capabilities. SeeSR \cite{lora_diff_realsr_wu2024seesr} enhances generative power via semantic prompts, while OSEDiff \cite{lora_diff_realsr_wu2024one} reduces diffusion steps by distilling text-to-image models into a one-step SR process. However, since both methods leverage costly diffusion models for SR, the efficiency gain from using LoRA becomes insignificant within the overall compute budget. Specifically, SeeSR \cite{lora_diff_realsr_wu2024seesr} uses LoRA to fine-tune a 750M-parameter prompt extractor over a week, and OSEDiff \cite{lora_diff_realsr_wu2024one} with 8.5M parameters uses LoRA to fine-tune its Stable Diffusion \cite{wang2024exploiting} backbone, both of which still suffer significantly in PSNR fidelity while handling real-world degradations. By contrast, our method leverages LoRA to efficiently adapt a bicubic-trained model to real SR domains with just 886k trainable parameters, achieving better perceptual and fidelity scores in one hour.

\noindent


\section{Methodology}
\label{method}

\noindent
\textbf{Preliminary: Low-Rank Adaptation (LoRA)} \cite{lora_orig_hu2021lora} fine-tunes models by updating a low-rank matrix approximation of weights, $\Delta W = BA$, where $W_0 + \Delta W = W_0 + (\alpha/r) BA$. Here, $W_0$ is the pre-trained matrix, and $B \in \mathbb{R}^{d \times r}$ and $A \in \mathbb{R}^{r \times k}$ are low-rank matrices, with rank $r \ll \min(d, k)$ and $\alpha$ is the scaling factor of updated weights. This setup freezes $W_0$ during training, updating only $A$ and $B$, making LoRA a highly efficient method for rapid adaptation with limited computational resources.

\subsection{Layer-Wise LoRA Integration}

\noindent
\textbf{Conv-LoRA Layers.} In Transformer/CNN-based SR architectures, convolution layers are essential for feature extraction and image reconstruction. In Transformers, convolution layers are employed for initial feature extraction, within residual blocks for high-frequency detail learning, and in final layers for SR output reconstruction. Conversely, CNNs depend entirely on convolutional layers for feature extraction and upsampling, effectively capturing essential texture and structural information for precise SR. Consider a convolutional layer with weights \( W \in \mathbb{R}^{C_{\text{out}} \times C_{\text{in}} \times k \times k} \), where \( C_{\text{out}} \) is the number of output and, \(C_{\text{in}} \) is the number of input channels, and \( k \times k \) is the spatial kernel size. The LoRA-modified convolution layer is defined as follows:


\begin{equation}
  \text{Conv}_{\text{LoRA}}(x) = (W + \alpha B_{\text{conv}} A_{\text{conv}}) * x
\end{equation}
\noindent
\( A_{\text{conv}} \in \mathbb{R}^{r \times C_{\text{in}} \times k \times k} \) and \( B_{\text{conv}} \in \mathbb{R}^{C_{\text{out}} \times r \times k \times k} \) are low-rank matrices (with \( r \) much smaller than \( C_{\text{out}} \) and \( C_{\text{in}} \)), \( \alpha \) is a scaling factor and \( * \) denotes the convolution operation.

\noindent
\textbf{MLP LoRA Layers.} In SR architectures, linear layers from the multi-layer perceptron (MLP) and attention blocks are pivotal for capturing global and contextual relationships across the image. Linear layers of MLP module help refinement of the feature representations learned by the attention mechanisms and enhance long-range dependencies. Therefore, we introduce a low-rank decomposition to the weight matrices of the MLP layer as:
\begin{equation}
 W_1' = W_1 + \alpha B_{lin_1}A_{lin_1} \quad
 W_2' = W_2 + \alpha B_{lin_2}A_{lin_2}
 \end{equation}

\noindent
where \( A_{lin_1} \) and \(A_{lin_2} \) \(\in \mathbb{R}^{r \times d} \) and \( B_{lin_1} \)  and \( B_{lin_2} \) \( \in \mathbb{R}^{d \times r} \) are learnable matrices with \( r \ll d \), \( \alpha \) is a scaling factor that controls the update magnitude \( W_1' \) and \( W_2' \) are the modified weight matrix with LoRA applied. Then, the MLP operation with LoRA becomes:
\begin{equation}
    \text{MLP}_{\text{LoRA}}(x) = W_2' \sigma\left(W_1' ( x ) \right)
\end{equation}

\noindent
where \( x \) is the input feature vector and \( \sigma \) is an activation function GELU \cite{gelu_hendrycks2016gaussian}.

\noindent
\textbf{MSA LoRA Layers.} 
In the context of multi-head self-attention (MSA) within Transformer-based SR models, attention layers are pivotal for capturing relationships between distant pixels, enhancing structural understanding and texture synthesis. LoRA is applied to the projection matrices of the query (\( W_Q \)), key (\( W_K \)), and value (\( W_V \)) in the attention mechanism of merged linear layers. The projections incorporating LoRA are formulated as:

\begin{multline}
   Q = x(W_Q + \Delta W_Q), \quad K = x(W_K + \Delta W_K), \\
   \quad V = x(W_V + \Delta W_V)
\end{multline}

where \( W_Q, W_K, W_V \in \mathbb{R}^{d \times d_k} \) are the weight matrices that project the input \( x \) to the query, key, and value representations, and  \( \Delta W_Q = B_Q A_Q \), \( \Delta W_K = B_K A_K \), and \( \Delta W_V = B_V A_V \), with \( A_Q, A_K, A_V \in \mathbb{R}^{r \times d} \) and \( B_Q, B_K, B_V \in \mathbb{R}^{d_k \times r} \), where \( r \ll d \) and \( d_k = \frac{d}{h} \) (with \( h \) being the number of attention heads). 

Given an input \( x \in \mathbb{R}^{N \times d} \), where \( N \) is the number of tokens and \( d \) is the embedding dimension, the multi-head self-attention mechanism computes the attention scores \( A \) with an added learned relative position bias \( B \):

\begin{equation}
    \resizebox{0.92\hsize}{!}{
       $A = \text{softmax} \left( \frac{(x(W_Q + \Delta W_Q))(x(W_K + \Delta W_K))^T}{\sqrt{d_k}} + B \right)$ 
        }
\end{equation}

The output of the attention mechanism is then obtained by:

\begin{equation}
    \text{Attention}_{\text{LoRA}}(Q, K, V) = AV
\end{equation}

By integrating LoRA into both convolutional and attention layers, our approach efficiently adapts pre-trained bicubic models to real-world SR tasks, significantly reducing the number of trainable parameters and computational overhead while maintaining high-quality reconstruction performance.

\subsection{AdaptSR: LoRA-driven Transformer SR}
We develop our proposed method AdaptSR using the widely adopted SwinIR \cite{liang2021swinir} architecture, known for its robust combination of transformer-based global feature extraction and localized context modeling, making it well-suited for high-quality SR across diverse degradations. To efficiently adapt a bicubic-trained SwinIR model to real-world degradations, we strategically integrate LoRA layers into key architectural components, as illustrated in Figure \ref{fig:arch}. Specifically, Conv-LoRA layers are applied in the shallow feature extraction stage to maintain parameter efficiency while preserving essential low-level features. In the deep feature extraction stage, we introduce six Residual Transformer LoRA Blocks (RTLB), each containing six Transformer LoRA Layers (TLL), embedding LoRA-enhanced attention mechanisms (Figure \ref{fig:msa}) to effectively model both local and global context with minimal computational overhead. Additionally, Linear-LoRA layers are integrated into the MLP stage to further optimize feature transformation, while extra Conv-LoRA layers are employed before and after the upsampling stage to ensure high-resolution outputs with minimal parameter overhead. During training, only the low-rank LoRA matrices (\( A_{\text{lin}} \), \( B_{\text{lin}} \) for linear layers and \( A_{\text{conv}} \), \( B_{\text{conv}} \) for convolution layers) are updated, while the original weights \( W \) remain frozen. This ensures efficient, memory-friendly adaptation without disrupting the model’s learned feature representations. By selectively applying LoRA to pre-trained layers, AdaptSR enables the model to refine its feature transformations dynamically, adjusting to less structured degradations in real-world SR while preserving inference efficiency.

\section{Experiments}
\subsection{Experimental Settings}
\noindent
\textbf{Datasets.}
We train AdaptSR on DIV2K \cite{Agustsson_2017_CVPR_Workshops} and RealSR \cite{realsr_cai2019toward} training sets, using the Real-ESRGAN \cite{wang2021real} degradation pipeline to create LR-HR pairs. During training, we also utilize the DIV2K unknown degradation dataset to cover a wider range of degradation types, using randomly cropped $256\times256$ patches from the HR images. For evaluation, we benchmark against other methods on DIV2K \cite{Agustsson_2017_CVPR_Workshops}, RealSR \cite{realsr_cai2019toward}, and DRealSR \cite{drealsr_wei2020component} test sets. Especially to compare with diffusion-based approaches, we adopt the StableSR \cite{wang2024exploiting} validation setup, including 3000 patches from DIV2K, 100 from RealSR, and 93 from DRealSR, with LR-HR pairs sized at $128\times128$ and $512\times512$.

\noindent
\textbf{Implementation Details and Evaluation.}
All experiments were run on an NVIDIA Quadro RTX 4090 GPU using PyTorch. LoRA models rank 8 with \( \alpha \) value 1 were trained for 100k iterations with Adam \cite{kingma2014adam} and L1 loss, starting with a learning rate of $1e^{-3}$, reduced by 25$\%$ at 50k, 75k, and 90k iterations. In comparison, full model fine-tuning required 500k iterations to achieve comparable results, with the learning rate halved at 250k, 400k, 450k and 475k intervals. Both LoRA and fine-tuning used the same settings, including the Adam optimizer \cite{kingma2014adam}, a batch size of 8, and the L1 loss function. To evaluate fidelity and perceptual quality, we used PSNR and SSIM (Y channel in YCbCr space) and perceptual metrics LPIPS \cite{lpips} and DISTS \cite{dists}.

\begin{table*}[t!]
 \caption{\textit{Quantitative comparison of our proposed AdaptSR with the state-of-the-art real-world GAN and diffusion methods}. Our approach trained only by L1 loss achieves high PSNR and SSIM scores while also delivering perceptual scores comparable to GAN and diffusion-based methods trained with perceptual losses. The best results are marked in \textbf{bold}.}
    \centering
    \scalebox{0.7}{
    \begin{tabular}{c|l|cccc|ccccccc|c}
    \specialrule{.1em}{.05em}{.05em} 
    & & \multicolumn{4}{|c|}{GAN-based Real SR} & \multicolumn{7}{c|}{Diffusion-based Real SR} & \multicolumn{1}{c}{} \\
    \hline
    & & RealESRGAN & LDL & DASR & Real-SAFMN-L & StableSR & DiffBIR & ResShift & SeeSR & PASD & SinSR & OSEDiff & AdaptSR \\
    & & \cite{wang2018esrgan} & \cite{liang2022details} & \cite{liang2022efficient} & \cite{safmn_sun2023spatially} & \cite{wang2024exploiting} & \cite{rebuttal_lin2025diffbir} & \cite{yue2023resshift} & \cite{lora_diff_realsr_wu2024seesr} & \cite{yang2023pasd} & \cite{wang2024sinsr} & \cite{lora_diff_realsr_wu2024one} & (Ours) \\ 
    \hline
     \multirow{2}{*}{\centering Dataset} & Trainable & \multirow{2}{*}{\centering 17M} & \multirow{2}{*}{\centering 12M} & \multirow{2}{*}{\centering M} & \multirow{2}{*}{\centering 5.6M}& \multirow{2}{*}{\centering 150M} & \multirow{2}{*}{\centering 380M} & \multirow{2}{*}{\centering 118.6M} & \multirow{2}{*}{\centering 750M} & \multirow{2}{*}{\centering 625M} & \multirow{2}{*}{\centering 119M}& \multirow{2}{*}{\centering 8.5M}& \multirow{2}{*}{\centering 886k} \\
    & Parameters & & & & & & & & & & & \\
    \hline
    \multirow{4}{*}{\centering DIV2K} & PSNR $\uparrow$ & 22.94 & 23.76 & 21.72 & 23.21 & 23.68 & 20.84 & 20.94 & 21.75 & 23.14 & 24.41 & 23.72 & \textbf{25.81} \\
    & SSIM $\uparrow$ &  0.6036 & 0.6403 &  0.5535 & 0.5950& 0.4887 & 0.4938 & 0.5422 & 0.6043 &  0.5505 &  0.6018 &  0.6108 & \textbf{0.6681}\\
    & LPIPS $\downarrow$ &  0.3768 & 0.3091 & 0.4266 &  0.3282& 0.4055 & 0.4270 & 0.4284 & 0.3194 &  0.3571 &  0.3240 &  \textbf{0.2941} & 0.5047 \\
    & DISTS $\downarrow$ &  0.2520&  0.2189& 0.2688 & 0.2153& 0.2542 & 0.2471 & 0.2606 & \textbf{0.1968}&  0.2207&  0.2066&  0.1976 & 0.3194\\
    \hline
    \multirow{4}{*}{\centering RealSR} & PSNR $\uparrow$ & 25.69 & 24.96 & 27.02 & 24.23 & 24.70 & 24.77 & 26.31 & 25.18 & 25.21 & 26.28 & 25.15 & 28.70\\
    & SSIM $\uparrow$ &  0.7616 & 0.7634 & 0.7708 & 0.7217& 0.7085 & 0.6572 & 0.7421 &  0.7216&  0.6798&  0.7347&  0.7341& \textbf{0.8079} \\
    & LPIPS $\downarrow$ &  0.2727 & \textbf{0.2519}& 0.3151 &  0.2905& 0.3018 & 0.3658 & 0.3460 & 0.3009&  0.3380&  0.3188&  0.2921& 0.2591\\
    & DISTS $\downarrow$ & 0.2063 & \textbf{0.1981}& 0.2207 &  0.2176& 0.2135 & 0.2310 & 0.2498 &  0.2223&  0.2260&  0.2353&  0.2128&  0.2109 \\
     \hline
    \multirow{4}{*}{\centering DRealSR} & PSNR $\uparrow$ & 28.64 & 27.43 & 29.77 & 27.15 & 28.13 & 26.76 & 28.46 & 28.17 & 27.36 & 28.36 & 27.92 & \textbf{30.74} \\
    & SSIM $\uparrow$ & 0.8053 & 0.8078 & 0.7572 & 0.7671& 0.7542 & 0.6576 & 0.7673 & 0.7691& 0.7073& 0.7515& 0.7835& \textbf{0.8422} \\
    & LPIPS $\downarrow$ & 0.2847 & \textbf{0.2655} & 0.3169 & 0.3148& 0.3315 & 0.4599 & 0.4006 & 0.3189& 0.3760& 0.3665& 0.2968& 0.3381 \\
    & DISTS $\downarrow$ & 0.2089 & \textbf{0.2055} & 0.2235 & 0.2219 & 0.2263 & 0.2749 & 0.2656 & 0.2315 & 0.2531 & 0.2485 & 0.2165 & 0.2524\\
\specialrule{.1em}{.05em}{.05em} 
    \end{tabular} 
}
\label{table:diff_arc_lora}
\end{table*}

\begin{figure*}[t!]
\centering
\begin{subfigure}{0.15\textwidth}
    \includegraphics[width=\textwidth]{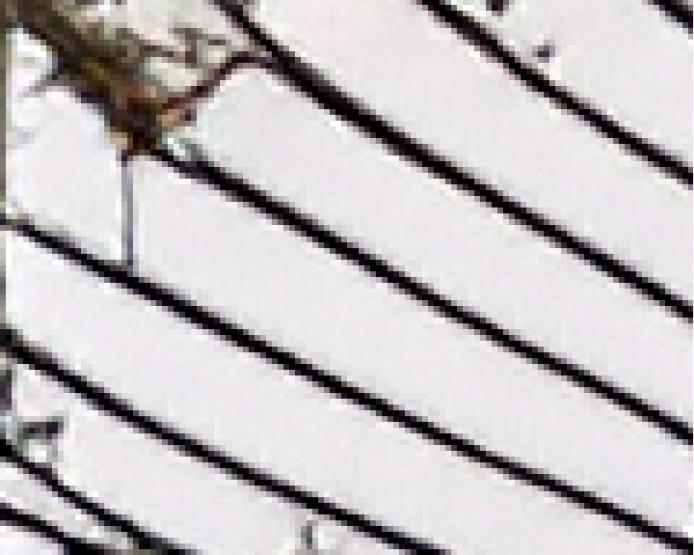}
    \vspace{-8pt}\\ \centering \scriptsize LDL  \\ \cite{liang2022details} \\ (9.627 / 0.727)
\end{subfigure}
\begin{subfigure}{0.15\textwidth}
    \includegraphics[width=\textwidth]{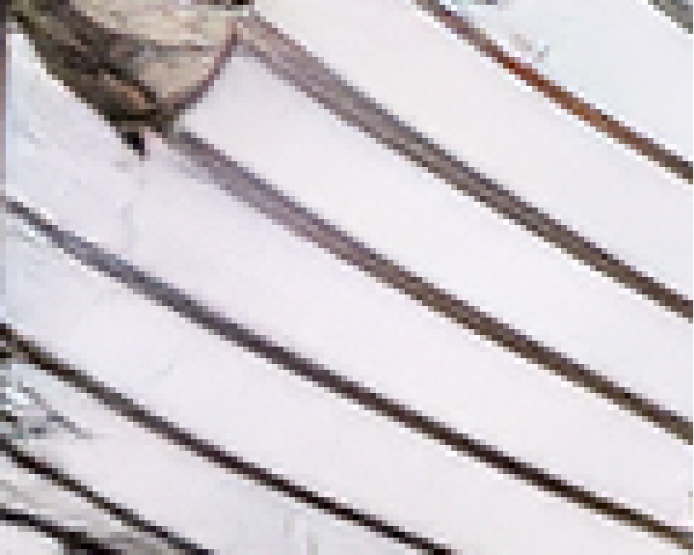} \vspace{-8pt}\\\centering \scriptsize  Real SAFMN-L \\ \cite{safmn_sun2023spatially} \\ (10.686 / 0.760)
\end{subfigure}
\begin{subfigure}{0.15\textwidth}
    \includegraphics[width=\textwidth]{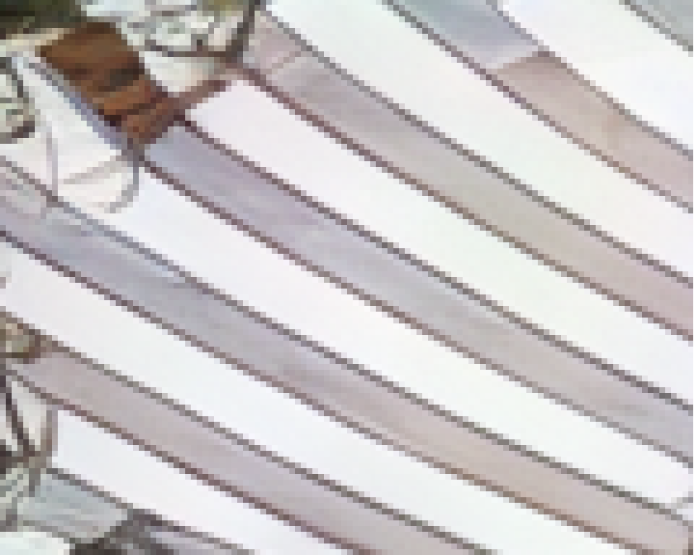} \vspace{-8pt}\\\centering \scriptsize PASD \\ \cite{yang2023pasd} \\ (11.276 / 0.768)
\end{subfigure}
\begin{subfigure}{0.15\textwidth}
    \includegraphics[width=\textwidth]{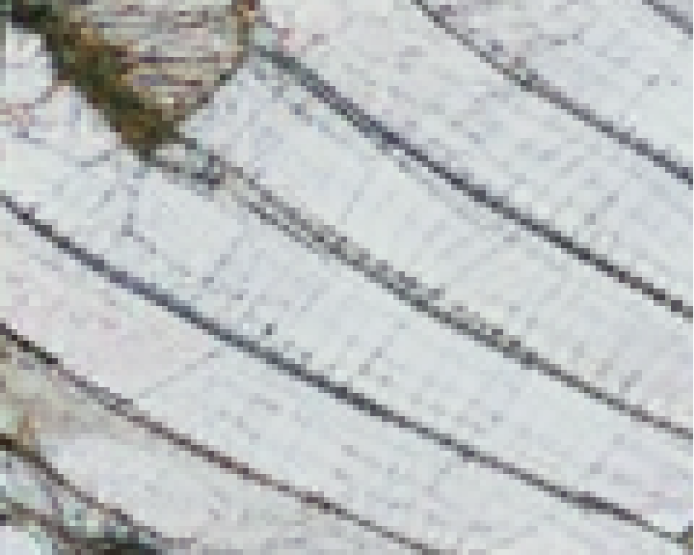} \vspace{-8pt}\\ \centering \scriptsize OSEDiff \\ \cite{lora_diff_realsr_wu2024one} \\ (11.559 / 0.772)
\end{subfigure}
\begin{subfigure}{0.15\textwidth}
    \includegraphics[width=\textwidth]{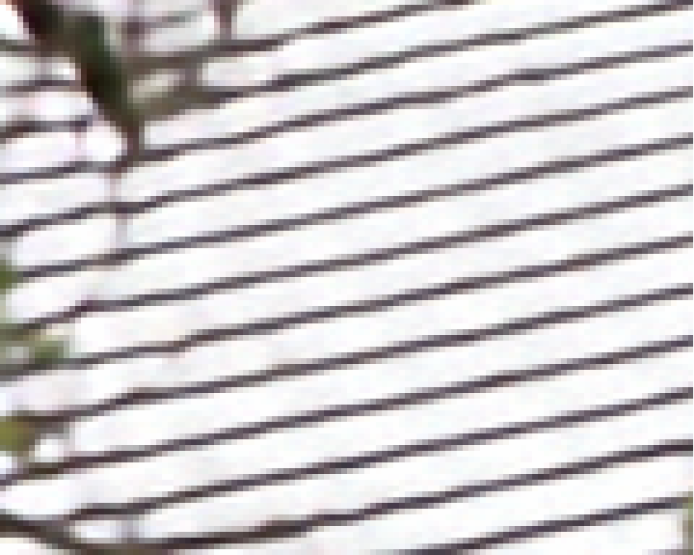} \vspace{-8pt}\\ \centering \scriptsize AdaptSR \\ (Ours) \\ (18.269 / 0.895)
\end{subfigure}
 \begin{subfigure}{0.15\textwidth}
    \includegraphics[width=\textwidth]{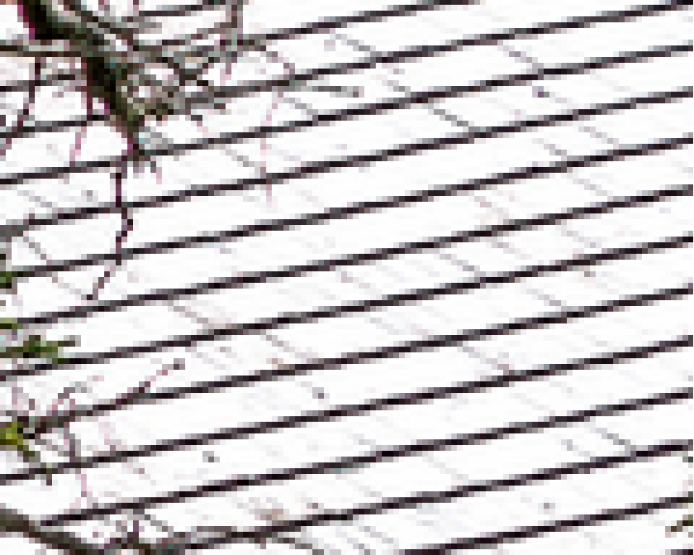} \vspace{-8pt}\\ \centering \scriptsize HR \\ Crop \\ (PSNR / SSIM)
\end{subfigure}

\begin{subfigure}{0.15\textwidth}
\includegraphics[width=\textwidth]{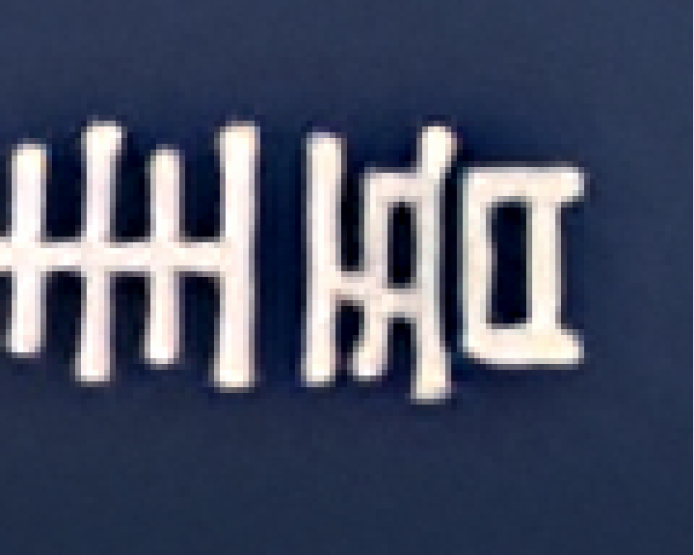}
\vspace{-8pt}\\ \centering \scriptsize LDL \\ \cite{liang2022details} \\ (19.119 / 0.936)
\end{subfigure}
\begin{subfigure}{0.15\textwidth}
\includegraphics[width=\textwidth]{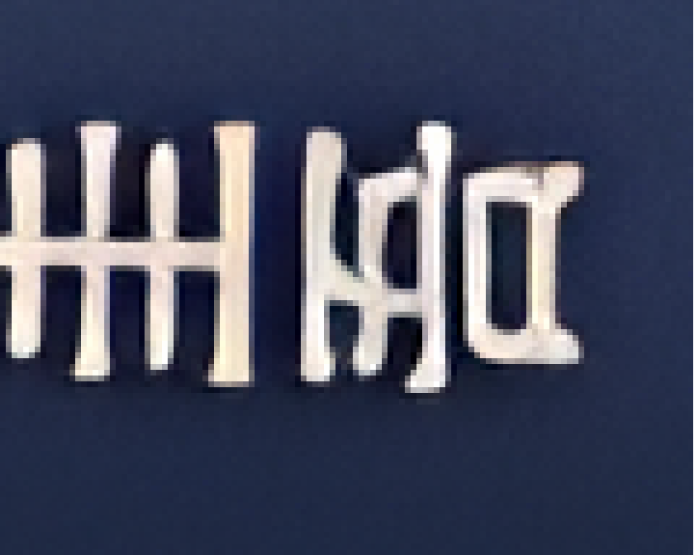} \vspace{-8pt}\\\centering \scriptsize  Real SAFMN-L \\ \cite{safmn_sun2023spatially} \\ (18.530 / 0.928)
\end{subfigure}
    \begin{subfigure}{0.15\textwidth}
    \includegraphics[width=\textwidth]{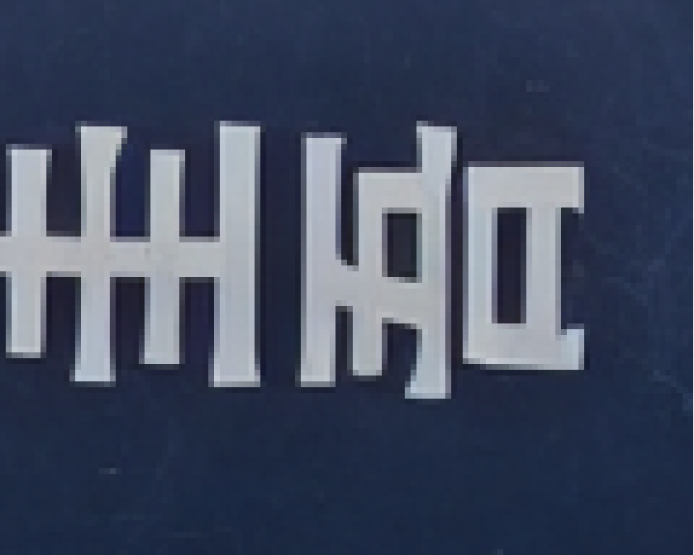} \vspace{-8pt}\\\centering \scriptsize PASD \\ \cite{yang2023pasd} \\ (21.608 / 0.953)
\end{subfigure}
\begin{subfigure}{0.15\textwidth}
    \includegraphics[width=\textwidth]{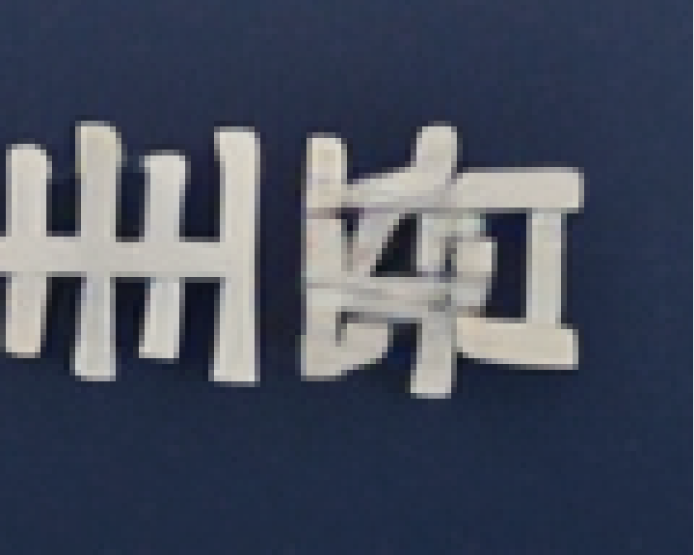} \vspace{-8pt}\\ \centering \scriptsize OSEDiff \\ \cite{lora_diff_realsr_wu2024one} \\ (19.801 / 0.943)
\end{subfigure}
\begin{subfigure}{0.15\textwidth}
    \includegraphics[width=\textwidth]{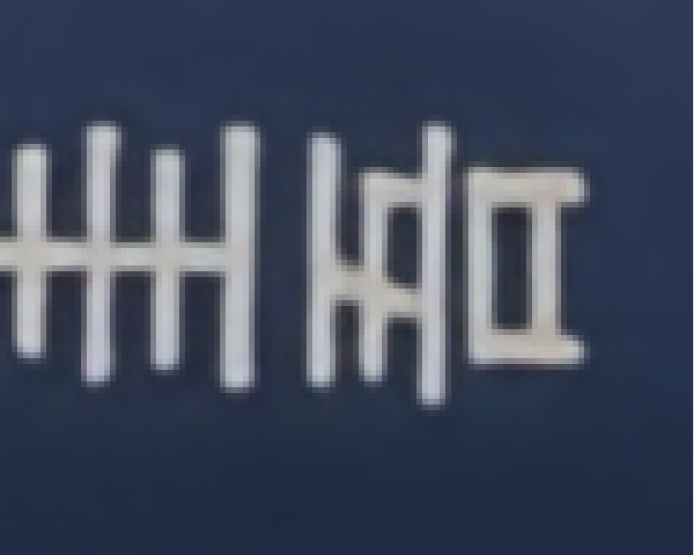} \vspace{-8pt}\\ \centering \scriptsize AdaptSR \\ (Ours) \\ (29.463 / 0.987)
\end{subfigure}
 \begin{subfigure}{0.15\textwidth}
    \includegraphics[width=\textwidth]{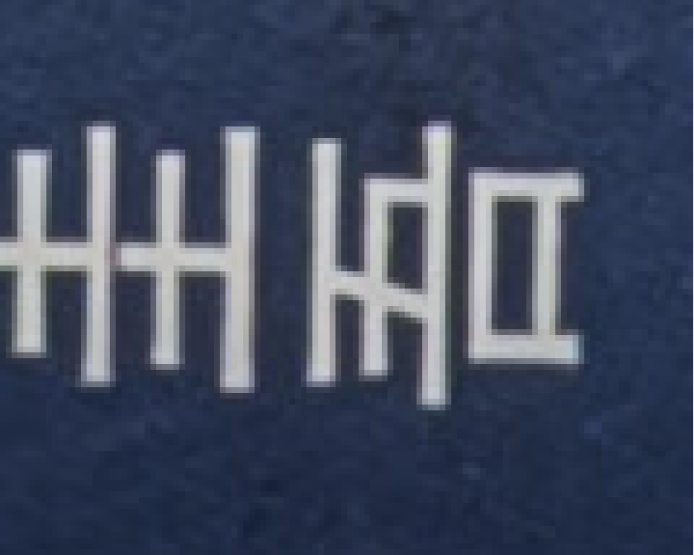} \vspace{-8pt}\\ \centering \scriptsize HR \\ Crop  \\ (PSNR / SSIM)
\end{subfigure}
\vspace{-3pt}
\caption{Visual comparison of the proposed AdaptSR with the state-of-the-art methods for $\times$4 real SR. GAN and diffusion models fail to capture the correct content of images, exhibit excessive sharpness with color shifts. On the other hand, our LoRA-based models reconstruct high-fidelity details with correct alignment, particularly in complex areas with regular patterns. \textit{Further visual comparisons are provided in the supplementary materials.}}
\label{fig:qual_results} 
\end{figure*}

\subsection{Comparison with State-of-the-Art}
\label{sec:comparison}
\textbf{Quantitative Performance.} Table \ref{table:diff_arc_lora} presents a quantitative comparison between our AdaptSR method and state-of-the-art real SR models, including GAN-based approaches (\textit{e.g.}, RealESRGAN \cite{wang2021real}, LDL \cite{liang2022details}) and diffusion methods (\textit{e.g.}, StableSR \cite{yang2023pasd}, DiffBIR \cite{rebuttal_lin2025diffbir}, ResShift \cite{yue2023resshift}, SeeSR \cite{lora_diff_realsr_wu2024seesr}, PASD \cite{yang2023pasd}, SinSR \cite{wang2024sinsr}, OSEDiff \cite{lora_diff_realsr_wu2024one}). The results are evaluated on $\times$4 real SR benchmarks. Despite using 20× fewer parameters than GAN-based methods like RealESRGAN \cite{wang2021real} and LDL \cite{liang2022details}, AdaptSR (e.g., SwinIR+LoRA) improves PSNR by up to 4 dB and LPIPS by 3$\%$ on RealSR \cite{realsr_cai2019toward} while updating only 886k parameters. Similarly, while diffusion models such as SeeSR \cite{lora_diff_realsr_wu2024seesr} and PASD \cite{yang2023pasd} require over 600M parameters, and OSEDiff \cite{lora_diff_realsr_wu2024one}—which is still 10× larger than our model—struggles to match our performance, AdaptSR achieves 2 dB higher PSNR on DIV2K \cite{Agustsson_2017_CVPR_Workshops} and surpasses OSEDiff by approximately 3.5 dB and 2.8 dB on the RealSR \cite{realsr_cai2019toward} and DRealSR \cite{drealsr_wei2020component} datasets, respectively. To conclude, our lightweight LoRA-based method consistently outperforms both GAN and diffusion-based SoTA approaches across distortion metrics (PSNR, SSIM) while achieving competitive perceptual scores (LPIPS, DISTS) with minimal parameter overhead. This establishes AdaptSR as an efficient and scalable solution for real-world SR, bridging the gap between synthetic and real training without excessive computational costs.

\noindent
\textbf{Qualitative Performance.} Figure \ref{fig:qual_results} illustrates visual comparisons among $\times$4 real SR methods, emphasizing the strengths of our AdaptSR approach. The qualitative results align closely with quantitative outcomes, revealing that GAN-SR models like LDL \cite{liang2022details} and Real SAFMN-Large \cite{safmn_sun2023spatially} show excessive sharpness with content inaccuracies such as misaligned striped patterns on roofs. Similarly, diffusion methods like PASD \cite{yang2023pasd} and OSEDiff \cite{lora_diff_realsr_wu2024one} introduce artifacts with inaccurate orientations for the first image. In contrast, our AdaptSR method successfully reconstructs fine details, accurately aligning roof stripes. Furthermore, in the second image, LDL \cite{liang2022details} and Real SAFM-L \cite{safmn_sun2023spatially} incorrectly render structure and colors. Likewise, diffusion methods PASD \cite{yang2023pasd} and OSEDiff \cite{lora_diff_realsr_wu2024one} produce inaccurate semantic generation. On the contrary, AdaptSR preserves the correct structure and colors in the second case. Overall, our approach effectively controls artifacts while maintaining image fidelity, producing photorealistic, high-quality results in domain adaptation from bicubic to real SR.

\begin{figure}
\centering
\begin{subfigure}{0.14\textwidth}
     \begin{subfigure}{\textwidth}
        \includegraphics[width=\textwidth]{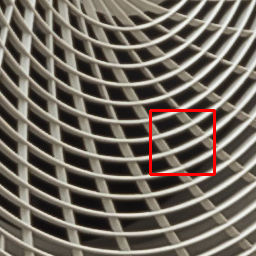} 
    \end{subfigure}
    \caption*{Nikon 41 \\ DI \cite{lam_gu2021interpreting}}
\end{subfigure}
\begin{subfigure}{0.14\textwidth}
     \begin{subfigure}{\textwidth}
        \includegraphics[width=\textwidth]{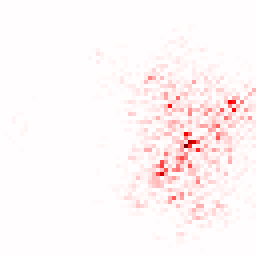} 
    \end{subfigure}
    \caption*{SwinIR \cite{liang2021swinir} FT \\  16.08}
\end{subfigure}
\begin{subfigure}{0.14\textwidth}
     \begin{subfigure}{\textwidth}
        \includegraphics[width=\textwidth]{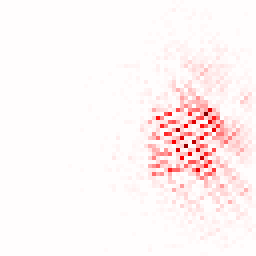} 
    \end{subfigure}
    \caption*{Attention Blocks \\ 14.72}
\end{subfigure}
\begin{subfigure}{0.14\textwidth}
    \begin{subfigure}{\textwidth}
        \includegraphics[width=\textwidth]{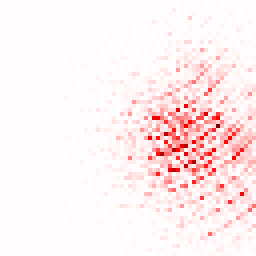} 
    \end{subfigure}
\caption*{MLP Blocks \\ 17.27}
\end{subfigure}
\begin{subfigure}{0.14\textwidth}
    \begin{subfigure}{\textwidth}
        \includegraphics[width=\textwidth]{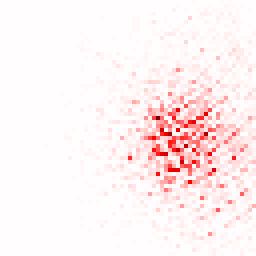} 
    \end{subfigure}
\caption*{Conv Layers \\ 18.78}
\end{subfigure}
\begin{subfigure}{0.14\textwidth}
    \begin{subfigure}{\textwidth}
        \includegraphics[width=\textwidth]{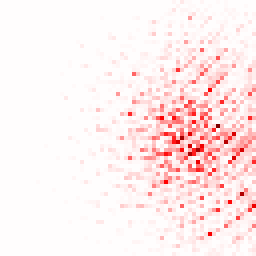} 
    \end{subfigure}
\caption*{All Layers \\ 19.59}
\end{subfigure}
\caption{Comparison of local attribution maps (LAMs) \cite{lam_gu2021interpreting} between different adaptation modules and standard FT model on RealSR-Nikon41  \cite{realsr_cai2019toward} image. The LAM results denote the importance of each pixel in the input image when super-resolving the patch marked with a red box. The diffusion index (DI) \cite{lam_gu2021interpreting} value reflects the range of involved pixels hence better reconstruction.} \vspace{-0.5cm}
\label{fig:saliency_image} 
\end{figure} 

\subsection{Adaptive Merging for Efficient Adaptation}
\label{sec:module-wise}
\begin{figure}
    \centering
    \includegraphics[width=0.45\textwidth]{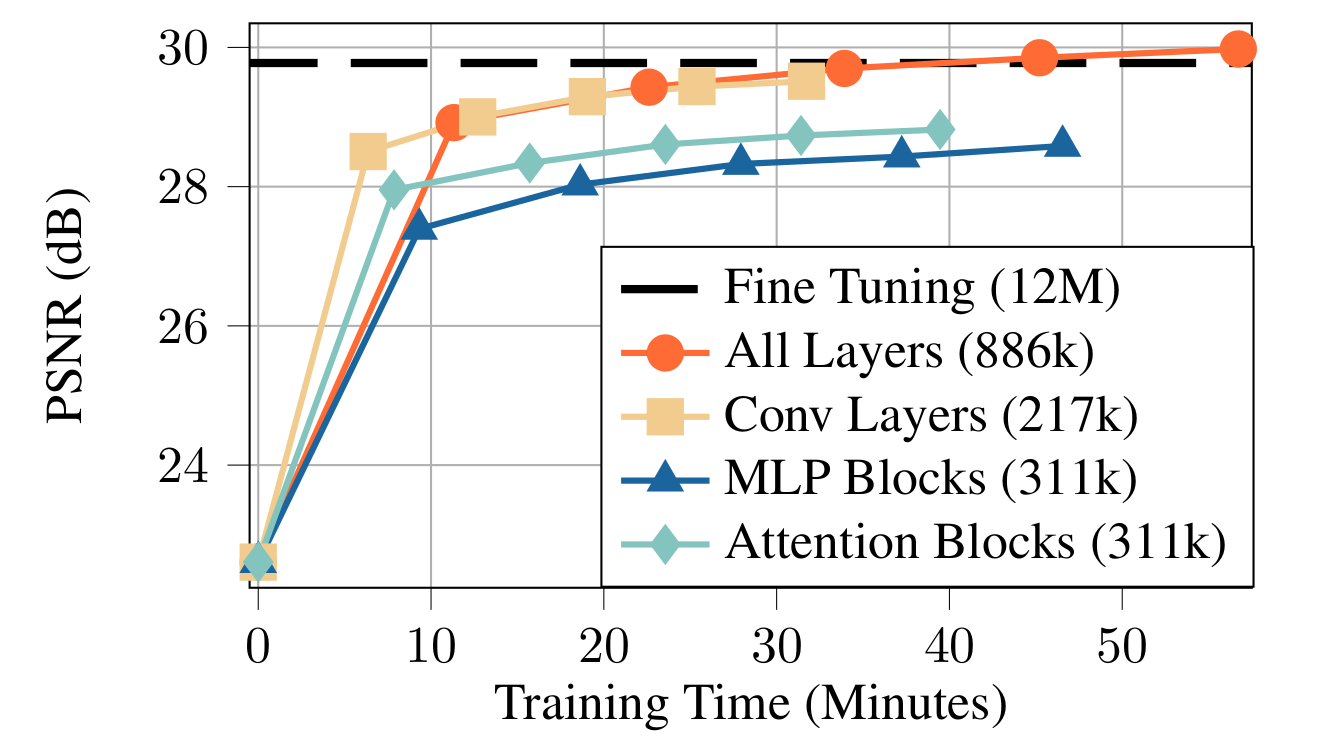} \vspace{-0.3cm}
\caption{Training time vs. performance comparison for different adaptation modules with 10k iterations on DIV2K unknown \cite{Agustsson_2017_CVPR_Workshops}. The dashed line indicates the best PSNR achieved by standard fine-tuning, which requires 4 days of training. Notably, LoRA applied to all layers reaches this PSNR level in just 30 minutes.}
\label{fig:training_time_vs_psnr}
\end{figure}

To determine the most critical modules for adapting Transformer architecture (i.e. SwinIR \cite{liang2021swinir}) from bicubic to real SR, we experimented with merging LoRA in four configurations: (1) all layers (convolutional, linear, and attention), (2) only convolutional layers, (3) only MSA layers, and (4) only MLP layers. As shown in Figure \ref{fig:training_time_vs_psnr}, applying LoRA to all layers achieved full fine-tuning performance in just 50 minutes (compared to 23 hours) while using only 8$\%$ of the parameters. Notably, LoRA applied exclusively to convolutional layers achieved nearly identical performance within 30 minutes, requiring just 2$\%$ of the parameters. Regardless of configuration, all methods started from
22.60 dB PSNR, matching the bicubic baseline.

To further analyze the contributions of different modules, we employed Local Attribution Maps (LAM) \cite{lam_gu2021interpreting}, which highlight feature interactions and contextual dependencies. As visualized in Figure \ref{fig:saliency_image}, applying LoRA across all layers enhances feature integration beyond individual module-based adaptations or full fine-tuning. Additionally, Diffusion Index (DI) \cite{lam_gu2021interpreting} values confirm that LoRA applied to all layers achieves the broadest attention range and strongest reconstruction quality, followed closely by the convolution-only configuration. Interestingly, even MLP-only and attention-only LoRA surpass standard fine-tuning in feature capture ability. This analysis demonstrates that LoRA applied to all layers is the optimal strategy for real SR adaptation, offering a balance between efficiency and reconstruction quality. However, for highly constrained settings, focusing solely on convolutional layers (2$\%$ parameters) provides an efficient yet competitive alternative.

\begin{figure}
\centering
\includegraphics[width=0.45\textwidth]{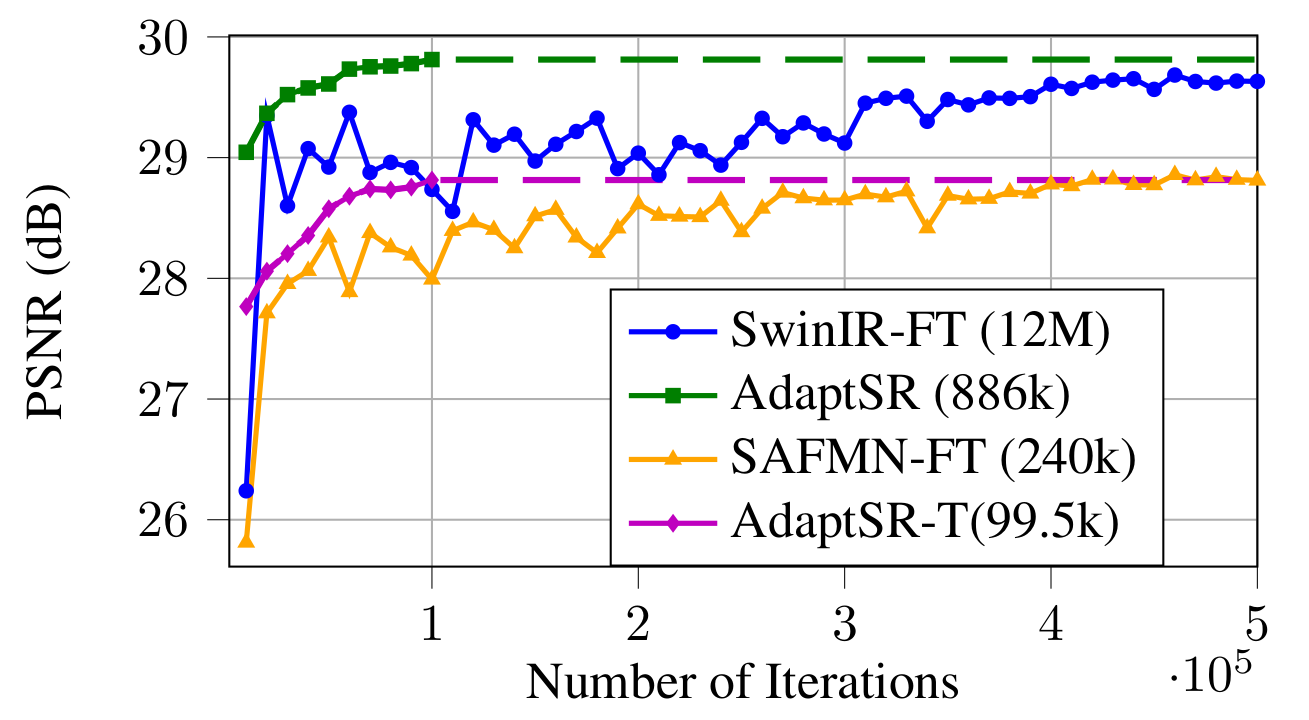}
 \vspace{-0.3cm}
\caption{Number of iterations vs. performance comparison for full model finetuning and our AdaptSR approach for SwinIR \cite{liang2021swinir} and SAFMN \cite{safmn_sun2023spatially}. LoRA reaches peak PSNR in just 100k iterations, completing in 4 hours for SwinIR and 1 hours for SAFMN on an NVIDIA RTX 4090, whereas fine-tuning takes significantly longer—23 hours for SwinIR and 9 hours for SAFMN—to achieve similar performance. The dashed lines of LoRA methods indicate their final PSNR performance at their latest iteration 100k.}
\label{fig:numiter_vs_psnr}
\end{figure}

\begin{table}
 \caption{\textit{Performance comparison for domain adaptation between fine-tuning and LoRA-enhanced models} on RealSR \cite{realsr_cai2019toward}. AdaptSR models match or outperform standard FT with significantly fewer trainable parameters and shorter training time, while maintaining the inference speed of baseline models. Here, $^{*}$ denotes the bicubic-trained model.}
    \centering
    \scalebox{0.7}{
    \begin{tabular}{c|cccccc}
    \specialrule{.1em}{.05em}{.05em} 
    \multirow{2}{*}{Model} & Trainable & FLOPs & GPU & \multirow{2}{*}{PSNR} & \multirow{2}{*}{SSIM} & \multirow{2}{*}{LPIPS}\\
     & Parameters & (G) & Hours &  & & \\  
    \hline
    EDSR$^{*}$ & 1.5M & - & - &27.5422 & 0.7732 & 0.3937\\
     +FineTuning & 1.5M & 64.97 & 10 & \textbf{28.0605}& \textbf{0.7963}& 0.3203 \\
     AdaptSR-C & 314k & 54.76 & 1.05 & 27.9260 & 0.7878 & \textbf{0.2835} \\
     \hdashline
    SAFMN$^{*}$ & 240k & - & - & 27.5383 & 0.7740 & 0.3927 \\
     + FineTuning & 240k & 7.75 & 9 &28.3173 & \textbf{0.7936 }& \textbf{0.2887}  \\
     AdaptSR-T & 99.5k & 5.89 & 1.17 & \textbf{28.3245} & 0.7935 & 0.2935  \\ 
     \hdashline
    SwinIR$^{*}$ & 12M  & - & - & 27.5537 & 0.7738 & 0.3965\\
     + FineTuning & 12M & 428.41 & 23 & 28.2537 & 0.7906 & 0.3820 \\
     AdaptSR & 886k & 410.57 & 4 & \textbf{28.6964} & \textbf{0.8079} &\textbf{ 0.2591}  \\ 
     \hdashline
    DRCT$^{*}$ & 14M & - & - & 27.5502 & 0.7737 & 0.3966  \\ 
     + FineTuning & 14M & 594.64 & 32 &\textbf{ 28.7481}& \textbf{0.8100}& 0.3007 \\
     + AdaptSR-L & 3M & 482.72 & 5.45 & 28.7173 & 0.8061 & \textbf{0.2553}\\
\specialrule{.1em}{.05em}{.05em} 
    \end{tabular} 
}
\label{table:diff_degredation_lora}
\end{table}

\subsection{AdaptSR vs. Fine-Tuning}
\label{sec:fine}
To comprehensively evaluate our LoRA-based domain adaptation approach, we applied it across diverse SR baselines. Table \ref{table:diff_degredation_lora} compares the performance of bicubic-trained models, their fine-tuned versions, and LoRA-based adaptations on the RealSR dataset \cite{realsr_cai2019toward}. We refer to our LoRA-based adaptation as AdaptSR when applied to SwinIR \cite{liang2021swinir}, AdaptSR-T (AdaptSR-Tiny) for lightweight SR models like SAFMN \cite{safmn_sun2023spatially}, AdaptSR-C (AdaptSR-CNN) for CNN-based models like EDSR \cite{EDSR2017}, and AdaptSR-L (AdaptSR-Large) for heavy transformer-based models like DRCT \cite{drct_Hsu_2024_CVPR}. Our results demonstrate that AdaptSR consistently matches or outperforms fine-tuning while drastically reducing trainable parameters, memory usage, and training time. For instance, in SwinIR, fine-tuning updates all 12M parameters, requiring 23 hours of training, whereas AdaptSR achieves superior performance with only 886k parameters and completes training in just 4 hours—a nearly 6× speedup. Specifically, AdaptSR improves PSNR by 0.42 dB, SSIM by 2$\%$, and LPIPS by 3.2$\%$ over fine-tuning, demonstrating both fidelity and perceptual quality gains. Similarly, in SAFMN, AdaptSR-T updates just 99.5k parameters, half the amount needed for fine-tuning, yet achieves the same PSNR improvement of 0.78 dB. The efficiency of AdaptSR becomes even more evident in larger models like DRCT. While FT requires 32 hours and 14M parameters, AdaptSR-L achieves comparable performance using just 3M parameters and 5.45 hours of training. Across all tested architectures, AdaptSR enhances both fidelity and perceptual quality, achieving up to a 1.14 dB PSNR increase and a 4.4$\%$ SSIM improvement over bicubic-trained models. Moreover, perceptual metrics such as LPIPS and DISTS confirm a reduction in artifacts and improved visual realism.

Another key advantage of AdaptSR is its rapid convergence. As shown in Figure \ref{fig:numiter_vs_psnr}, AdaptSR reaches peak performance within 100k iterations—approximately 4 hours for SwinIR and 1 hour for SAFMN on an NVIDIA RTX 4090—compared to 23 and 9 hours, respectively, for fine-tuning. Importantly, since LoRA layers merge into the base model post-training, the inference speed remains unchanged, making AdaptSR a practical and highly efficient alternative to fine-tuning for domain adaptation.

\subsection{Discussion}
\label{sec:discussion}

\noindent
\textbf{Comparison with Other PEFT Methods.}
\label{ss:comparison_peft}
Table \ref{table:adapter_vs_lora} compares AdaptSR (LoRA-based) with other PEFT methods on RealSR \cite{realsr_cai2019toward}. AdaptSR achieves a PSNR of 28.40 dB, surpassing most ARC variants and DiffFit while using fewer trainable parameters than ARC (svd500) and standard fine-tuning. Unlike ARC, which introduces additional inference overhead, AdaptSR seamlessly integrates into the base model post-training, preserving the original inference efficiency. Furthermore, AdaptSR consistently delivers strong perceptual quality, achieving the lowest LPIPS score, making it a more efficient and effective alternative for real-world SR adaptation.

\begin{table}
 \caption{\textit{PEFT methods vs. AdaptSR} comparison on RealSR \cite{realsr_cai2019toward}.} 
  \vspace{-0.3cm}
    \centering
    \scalebox{0.66}{
    \begin{tabular}{c|cccccl}
    \specialrule{.1em}{.05em}{.05em} 
    \multirow{2}{*}{Model} & \multirow{2}{*}{PSNR} & \multirow{2}{*}{SSIM} & \multirow{2}{*}{LPIPS} & \multirow{2}{*}{DISTS} & Updated / Inference Time\\ 
    & & & & & Parameters \\
    \hline
    Baseline & 27.5537& 0.7738& 0.3965& 0.2426 & - / 12M \\
    FT & 28.2637& 0.7922& 0.3016& 0.2274 & +12M / +12M \\
    ARC \cite{rebuttal_arc_dong2024efficient} (svd50) & 28.0836& 0.7894& 0.2995& 0.2206 & +125k / +125k \\
    ARC  (svd100) & 28.0840& 0.7933& 0.2883& 0.2189 & +236k \ +236k \\
    ARC (svd350) & 28.2216& 0.7954& 0.2735& \textbf{0.2127} & +794k \ +794k\\
    ARC (svd375) & 28.3657& 0\textbf{.7993}& 0.2714& 0.2129 & +850k / +850k \\
    ARC (svd400) & 28.0856& 0.7924& 0.2760& 0.2144  & +906k / +906k \\ 
    ARC (svd500) & 28.1611& 0.7949& 0.2711& 0.2130 & +1.13M / +1.13M \\
    DiffFit \cite{rebuttal_xie2023difffit} & 28.2695& 0.7949& 0.2949& 0.2199& +13k / +13k \\
    AdaptSR (Ours) & \textbf{28.3999}& 0.7976 & \textbf{0.2705}& 0.2128 & +886k / -\\ 
    
\specialrule{.1em}{.05em}{.05em} 
    \end{tabular} 
}
\label{table:adapter_vs_lora}
\vspace{-0.22cm}
\end{table}

\begin{table}
 \caption{\textit{Module-wise comparison between LoRA-based AdaptSR and FT} for domain adaptation on DRealSR \cite{drealsr_wei2020component}.}
    \centering
    \scalebox{0.8}{
    \begin{tabular}{l|cc|cc|cc}
    \specialrule{.1em}{.05em}{.05em} 
    & \multicolumn{2}{c|}{Convs} &\multicolumn{2}{c|}{MLPs} &\multicolumn{2}{c}{MSAs} \\
    \hline
    & FT & LoRA &  FT & LoRA &  FT & LoRA \\
    \hline
    Params. & 2.2M & 868k & 4.7M & 1.2M &  4.7M & 1.2M \\
    PSNR & 30.29 & \textbf{30.97} & 30.82 & \textbf{31.02} & \textbf{30.83} & 30.77 \\   
    LPIPS & 0.3955 & \textbf{0.3924} & 0.3457 & \textbf{0.3399} & \textbf{0.3465} & 0.4023 \\
\specialrule{.1em}{.05em}{.05em} 
    \end{tabular} 
}
\label{table:conv_comparison}
\end{table}

\noindent
\textbf{Module-wise Comparison of AdaptSR vs. FT.}
Table \ref{table:conv_comparison} presents a layer-wise comparison between FT and AdaptSR on DRealSR \cite{drealsr_wei2020component} across three key modules: convolutional layers (Convs), MLPs, and MSAs. The results highlight AdaptSR’s efficiency in parameter reduction while maintaining or surpassing FT performance. AdaptSR achieves a higher PSNR in convolutional layers while updating only 868k parameters instead of 2.2M, demonstrating its efficiency in capturing low-level textures. In MLPs, it improves PSNR with 1.2M parameters (vs. 4.7M for fine-tuning), though at a slight perceptual cost. For MSAs, AdaptSR maintains comparable performance with far fewer parameters, but its efficiency gains are most significant in convolutional and MLP layers. Overall, AdaptSR provides a compact, computationally efficient alternative to full fine-tuning, particularly excelling in convolutional and MLP layers, where it matches or outperforms fine-tuning with significantly fewer parameters. These insights guide efficient adaptation strategies for transformer-based SR models.

\noindent
\textbf{Extention to GAN-Based Real SR.} We further extend LoRA-based domain adaptation to GAN-based SR with AdaptSR-GAN, demonstrating its flexibility and efficiency. Using SwinIR \cite{liang2021swinir} with LoRA layers and a U-Net discriminator with spectral normalization \cite{wang2021real}, we employ adversarial training with a weighted combination of L1, perceptual \cite{johnson2016perceptual}, and GAN losses (1:1:0.1) as in RealESRGAN. As shown in Table \ref{table:gan_result}, AdaptSR-GAN outperforms RealESRGAN \cite{wang2021real} and LDL \cite{liang2022details} in both fidelity and perceptual quality while requiring only 886k parameters—far fewer than the 12M in LDL and 17M in RealESRGAN. These results establish AdaptSR-GAN as an efficient, low-overhead alternative for GAN-based SR, effectively handling real-world degradations.

\begin{table}[t!]
 \caption{\textit{Comparison of GAN-based RealSR }methods and our adversarially trained AdaptSR-GAN on DIV2K \cite{Agustsson_2017_CVPR_Workshops} validation patches. Our method achieves superior fidelity and perceptual quality with significantly fewer parameters.}
    \centering
    \scalebox{0.8}{
    \begin{tabular}{cccc}
    \specialrule{.1em}{.05em}{.05em} 
      & LDL (RealSwinIR) & RealESRGAN & AdaptSR-GAN \\ 
      \hline
      Params. & 17M & 12M & 886k \\
    \hline
    PSNR $\uparrow$ & 23.76 & 21.94 & \textbf{24.11}\\
    SSIM $\uparrow$ & 0.6403 & 0.5736 & \textbf{0.6598 }\\
    LPIPS $\downarrow$ & 0.3091 & 0.3868 & \textbf{0.2914 }\\
    DISTS $\downarrow$ & 0.2189 & 0.2601 &\textbf{ 0.2185}  \\ 
\specialrule{.1em}{.05em}{.05em} 
    \end{tabular} 
}
\label{table:gan_result}
\end{table}

\subsection{Ablation Study}
\noindent
\textbf{Effect of CNN-LoRA Layers.} Table \ref{table:effect_of_lora_layers} shows the impact of layer-specific updates for adapting SwinIR \cite{liang2021swinir} to real SR. Partitioning and evaluating convolutional layers reveals that certain layers, especially those in residual blocks (RSTLB Conv) and the final convolution layer after upsampling (AU Conv), achieve higher fidelity and perceptual scores with fewer parameter updates than MSA and MLP layers on DRealSR \cite{drealsr_wei2020component} at rank 32. Notably, the AU Conv layer delivers the highest performance gains, outperforming both the initial convolution layer (First Conv), the ones in the deep feature extraction module (DFE Convs) and pre-upsampling layer (BU Conv). Overall, adapting MLP linear layers or AU Conv optimally supports domain adaptation. 

\noindent
\textbf{Rank Analysis.}
\label{sec:rank}
To identify the optimal LoRA rank for real SR adaptation, we conduct an ablation study by incrementally testing ranks from \( r = 1 \) to 64. Performance on the RealSR \cite{realsr_cai2019toward} dataset for $\times$4 images, shown in Table \ref{table:rank_analysis}, demonstrates that even at rank 1 (with only 1.4$\%$ of the trainable parameters), the results closely match those from fine-tuning with 12M parameters for SwinIR \cite{liang2021swinir}. Optimal fidelity for PSNR and SSIM and perceptual quality for LPIPS are achieved at \( r = 8 \). Higher ranks (16 and 64) do not yield additional gains but still outperform full fine-tuning with fewer parameter updates. These results align with prior findings of \cite{zeng2023expressive}, suggesting that larger models often require lower LoRA ranks for effective adaptation, especially when pre-trained models are well-suited to the target task. \textit{Ablation studies for scaling factor are in supplementary materials.}

\begin{table}
 \caption{\textit{Effect of CNN-LoRA layers for domain adaptation} from bicubic to real SR on DRealSR \cite{drealsr_wei2020component} dataset for rank 32.}
    \centering
    \scalebox{0.75}{
    \begin{tabular}{cccccc}
    \specialrule{.1em}{.05em}{.05em} 
    \multirow{2}{*}{LoRA Layer} & \multirow{2}{*}{PSNR} & \multirow{2}{*}{SSIM} & \multirow{2}{*}{LPIPS} & Trainable &  Training \\
    & & & & Parameters & Time (Mins)\\ 
    \hline
   MSAs  & 30.771 & 0.8402 & 0.4023 & 1.2M & 77 \\
   MLPs  & \textbf{31.024} & 0.8433 & 0.3999 & 1.2M & 57.71 \\
    All Convs & 30.974 & \textbf{0.8425} & \textbf{0.3994} & 868k & 31.45 \\
   \hdashline
   First Conv & 30.078 & 0.8226 & 0.4327 & 53k & 24.95 \\
   RSTLB Convs & 30.968 & 0.8422 & 0.4003 & 622k & 41.09 \\
   DFE Convs & 29.530 & 0.8088 & 0.4470 & 104k & 19.22 \\
   BU Conv & 29.583 & 0.8117 & 0.4379 & 70k & 19.02  \\
   AU Conv & 30.479 & 0.8305 & 0.4201 & 19k & 18.42 \\
\specialrule{.1em}{.05em}{.05em} 
    \end{tabular} 
}
\label{table:effect_of_lora_layers}
\end{table}

\begin{table}
 \caption{\textit{Rank analysis} for better domain adaptation on RealSR \cite{realsr_cai2019toward} for 10k iterations. The best result is marked in \textbf{bold}.}
    \centering
    \scalebox{0.72}{
    \begin{tabular}{ccccccl}
    \specialrule{.1em}{.05em}{.05em} 
    Model & PSNR & SSIM & LPIPS & DISTS & Updated Paramaters\\  
    \hline
    Baseline & 27.5537& 0.7738& 0.3965& 0.2426 & 12M \\
    Baseline +FT & 28.2637& 0.7922& 0.3016& 0.2274 & +12M \\
    \hdashline
    r=1 &28.3383& 0.7957& 0.2772& 0.2129 &+152k (1.4\%)\\ 
    r=4 & 28.2675& 0.7945& 0.2721& \textbf{0.2125} & +466k (4.2\%)\\ 
    r=8 & \textbf{28.3999}& \textbf{0.7976}& \textbf{0.2705}& 0.2128 & +886k (8.1\%)\\ 
    r=16 & 28.3575 & 0.7959 & 0.2719& 0.2140& +1.7M (15.5\%)\\ 
    r=64 & 28.3537& 0.7959 & 0.2716 & 0.2130 & +6.8M (61.8\%)\\ 
\specialrule{.1em}{.05em}{.05em} 
    \end{tabular}  \vspace{-0.5cm}
}
\label{table:rank_analysis}
\end{table}

\section{Conclusion}
We propose AdaptSR, a LoRA-based approach for efficient domain adaptation in SR models, bridging the gap between synthetic and real-world SR applications. Domain adaptation is a crucial challenge in SR, as strong bicubic models struggle with real-world degradations. Our method mitigates this failure, making real-world deployment feasible under low-memory and low-FLOP constraints. By updating only LoRA layers while preserving the pretrained bicubic model weights, our approach effectively leverages learned image priors to achieve comparable or superior performance to full fine-tuning at a fraction of the memory and compute cost—often within minutes. Unlike a trivial application of LoRA, our method is carefully designed for real SR adaptation through architecture-specific insights, selective layer updates, and extensive empirical validation across both CNN- and transformer-based SR architectures. This strategic optimization enables AdaptSR to outperform state-of-the-art GAN and diffusion models while reducing artifacts and maintaining high-quality outputs. The minimized updates in LoRA layers make our method stable, efficient, and highly adaptable, addressing memory and runtime constraints in real-world SR applications and making it a practical and scalable solution for real SR deployment.

\vspace{10pt}


\clearpage
\newpage

{
    \small
    \bibliographystyle{ieeenat_fullname}
    \bibliography{source}
}


\end{document}